\documentclass[aps,pre,twocolumn,groupedaddress,longbibliography]{revtex4-1}
\usepackage{graphicx,amsmath,amssymb,color}

\begin{document}

\title{Homogeneous crystallization in four-dimensional Lennard-Jones liquids}
\author{Robert S. Hoy}
\email{rshoy@usf.edu}
\affiliation{Department of Physics, University of South Florida, Tampa, FL 33620}
\date{\today}
\begin{abstract}
We report the first observation of homogeneous crystallization in simulated high-dimensional ($d > 3$) liquids that follow physically realistic dynamics and have system sizes that are large enough to eliminate the possibility that crystallization was induced by the periodic boundary conditions.
Supercooled four-dimensional (4D) Lennard-Jones liquids maintained at zero pressure and constant temperatures $0.59 < T < 0.63$ crystallized within $\sim 2 \times 10^4\tau$, where $\tau$ is the LJ time unit.
WCA liquids that were maintained at the same densities and temperatures at which their LJ counterparts nucleated did not crystallize even after $2.5\times 10^5\tau$, showing that the presence of long-ranged attractive interactions dramatically speeds up 4D crystallization, much as it does in 3D.
However, the overlap of the liquid and crystalline phases' local-bond-order distributions is much smaller for LJ than for WCA systems, which is the opposite of the 3D trend.
This implies that the widely accepted hypothesis that increasing geometrical frustration rapidly suppresses crystallization as the spatial dimension $d$ increases is only generally valid in the absence of attractive interparticle forces.
\end{abstract}
\maketitle

\section{Introduction}
\label{sec:Intro}

In three-dimensional liquids composed of monodisperse particles lacking strongly directional interactions, local structural ordering at the atomic scale is typically approximately icosahedral \cite{frank52}.
Icosahedra, which are composed of a central atom surrounded by 12 atoms that form a locally-fivefold-symmetric shell, are the lowest-energy 13-atom structures for a wide-range of pair potentials \cite{doye95}; they are composed of 20 distinct tetrahedra, which are the lowest-energy 4-atom structures.
Both tetrahedra and icosahedra, however, are incompatible with these potentials' lowest-energy \textit{global} structures, i.e.\ with the FCC and HCP crystal lattices.
Such incompatibility between the lowest-energy \textit{local} structures and the lowest-energy \textit{global} structures is known as ``geometrical frustration'' and is one of the best-known reasons for glass formation \cite{nelson83}.
It makes the free energy barriers for rearrangements from locally into globally preferred structures large, increasing atoms' tendency to stay in the former under rapid cooling or compression, and promoting the formation of locally polytetrahedral amorphous order at the expense of  close-packed crystalline order in a wide variety of 3D glassy and jammed solids \cite{anikeenko07}.

Three-dimensional systems are, however, somewhat unusual in this respect.
In two dimensions, the lowest-energy 7-atom clusters (for the same pair potentials that give icosahedra in 3D) are hexagons composed of 6 triangles, which are the lowest-energy 3-atom structures.
Since these structures are both compatible with the triangular lattice,  monodisperse 2D soft-sphere and hard-sphere liquids readily crystallize \cite{quasiLR} under a wide range of preparation protocols \cite{pieranski80,lubachevsky91,reis06,gonzalez16}.
In four dimensions, the lowest-energy/most-compact 25-atom clusters are composed of 24 octahedral cells, which are the lowest-energy/most-compact 8-atom structures.
Both of these structures are compatible with the $D_4$ lattice, which is  the densest 4D sphere packing \cite{conway93}.
Thus it was surprising to find that crystallization of 4D hard-sphere liquids is strongly suppressed.
In the first study of these liquids' solidification dynamics \cite{skoge06}, crystallization occurred in a very-slowly-compressed, very small (648-atom) system, but none was observed in a 10000-atom system, suggesting that the 648-atom result may have been a finite-size artifact arising from the periodic boundary conditions \cite{foot648}.
More recent studies of larger systems also failed to observe homogeneous crystallization \cite{vanMeel09,vanMeel09b,charbonneau10,charbonneau12,charbonneau13,charbonneau21,charbonneau21b,lue10,biasedMC}.

van Meel \textit{et.\ al} explained this failure in terms  of a less-obvious kind of geometrical frustration, namely that the \textit{actual} local ordering in equilibrated 4D hard-sphere liquids is very different than that of the abovementioned 25-atom clusters \cite{vanMeel09,vanMeel09b}.
More specifically, the overlap $\mathcal{O}$ of the probability distributions for  local bond order ($q_6$ \cite{steinhardt83}) in the metastable supercooled liquid and equilibrium crystalline states, 
\begin{equation}
\mathcal{O} = \displaystyle\int_0^{1}  P_{\rm liquid}(q_6) P_{\rm cryst}(q_6) dq_6,
\label{eq:OQ6}
\end{equation}
is far smaller in 4D than it is in 3D, and consequently the free energy barriers to crystal nucleation (and specifically, the interfacial free energy) are much higher. 
This type of entropically-driven frustration gets more dramatic  as the spatial dimension $d$ increases \cite{vanMeel09b}, consistent with the now-widely-accepted notion that crystallization rapidly gets harder with increasing $d$ \cite{skoge06}.
Its presence can explain why crystallization is suppressed in high $d$ despite the fact that systems' equilibrium freezing densities $\phi_{\rm f}$ remain well below their glass transition densities \cite{charbonneau21,charbonneau21b}.

On the other hand, there are reasons to question whether results for hard-sphere liquids can be generalized to other systems.
While the structure of liquids at the level of the pair correlation function $g(r)$ is determined almost completely by the repulsive-core part of the interatomic interactions (for systems maintained at fixed density $\rho$ and temperature $T$ \cite{weeks71}), longer-ranged attractive forces can exert a substantial influence on higher-order structural metrics that influence crystallization propensity \cite{taffs10}. 
Toxvaerd recently showed \cite{toxvaerd21} that increasing the potential cutoff radius $r_{\rm c}$ in 3D Lennard-Jones liquids from its WCA value ($r_{\rm c}^{\rm WCA} = 2^{1/6}\sigma$, where $\sigma$ is the LJ length unit) to $3.5\sigma$ (a value that produces attractive forces for particles in atoms' first, second, and third coordination shells) increased  both their $\mathcal{O}$ values and their nucleation rates by at least one order of magnitude.
This result leads naturally to the question:\ 
is the same combination of trends also present in higher $d$?

To the best of our knowledge, only three previously published particle-based simulation studies of liquids in $d > 3$ have included attractive interactions \cite{hloucha99,bruning09,sengupta13}, and none of these examined systems that could be expected to crystallize.
Here, using large-scale molecular dynamics simulations of 4D WCA/Lennard-Jones liquids, we show that the answer to the above question is, surprisingly, ``no''.
LJ liquids reproducibly nucleate and form high-quality $D_4$ crystals over times as small as $\sim 10^4\tau$ at densities and temperatures for which WCA liquids do not crystallize on any currently computationally feasible time scale.
This difference occurs despite the fact that the LJ systems have $\mathcal{O}$ values that are several orders of magnitude \textit{lower} than their WCA counterparts.
Our results imply that the widely accepted hypothesis that increasing geometrical frustration rapidly suppresses crystallization as the spatial dimension $d$ increases is only generally valid \textit{in the absence of attractive interparticle forces.}

\section{Methods}
\label{sec:methods}

All simulations were performed using \texttt{hdMD} \cite{hoy22}. 
Systems are composed of $N = 5\times 10^5$ particles of mass $m$, interacting via the truncated and shifted Lennard-Jones potential $U_\textrm{LJ}(r) = 4\epsilon[(\sigma/r)^{12} - (\sigma/r)^{6} - (\sigma/r_{c})^{12} + (\sigma/r_c)^{6}]$, where $\epsilon$ is the interparticle binding energy and $r_c$   is the cutoff radius.
Newton's equations of motion are integrated with a timestep $dt = \tau/125$, where $\tau = \sqrt{m\sigma^2/\epsilon}$ is the LJ time unit.
Periodic boundary conditions are applied along all four directions of hypercubic simulation cells.
After initially placing the particles randomly within the cells and minimizing their energy to reduce interparticle overlap, short NVT-ensemble equilibration runs are performed.
For the LJ ($r_c = 2.5\sigma$) systems, these are followed by long NPT-ensemble runs of length up to $10^5\tau$, where pressure is maintained at zero and temperature is held constant using a Berendsen thermo/barostat \cite{berendsen84}.
In the later stages of our study, we also performed NVT simulations of WCA ($r_c = 2^{1/6}\sigma$) liquids and both WCA and LJ $D_4$ single crystals; these will be described further in Section \ref{sec:results}.

We monitor multiple thermodynamic metrics such as the average pair interaction energy per particle $E_{\rm pair}$, particle number density $\rho$, and pair correlation function $g(r)$. 
We also monitor a structural order parameter that has been shown to effectively characterize the type of 4D crystallization we expect to encounter: specifically,  the second-order two-particle bond-order correlator \cite{vanMeel09,vanMeel09b}
\begin{equation}
q_6(i,j) =  \displaystyle\frac{1}{\mathcal{N}(i)\mathcal{N}(j)} \displaystyle\sum_{\alpha = 1}^{\mathcal{N}(i)}  \displaystyle\sum_{\beta = 1}^{\mathcal{N}(j)} G_6^1(\hat{r}_{i\alpha}\cdot\hat{r}_{j\beta}).
\label{eq:q6ijdef}
\end{equation}
We calculate $q_6(i,j)$ for all neighboring particles $i$ and $j$ that lie within each other's first coordination shells as defined by the first minimum of $g(r)$, i.e.\ all particle pairs ($i,j$) whose distance $r_{ij} = |\vec{r}_j - \vec{r}_i| < 1.4\sigma$. 
The sums in Eq.\ \ref{eq:q6ijdef} are performed over the $\mathcal{N}(i)$ neighbors of particle $i$ and $\mathcal{N}(j)$ neighbors of particle $j$ satisfying $r_{i\alpha}, r_{j\beta} < 1.4\sigma$.
$G_6^1$ is the Gegenbauer polynomial defined by \cite{gegenwiki}
\begin{equation}
G_6^1(x) = \displaystyle\sum_{k = 0}^3 \displaystyle\frac{(-1)^k (6 - k)!(2x)^{6-2k}}{k!(6-2k)!}.
\label{eq:G61}
\end{equation}
This rotationally-invariant correlator is defined such that $q_6(i,j) = 1$ in a perfect $D_4$ lattice and $|q_6(i,j)| \ll 1$ in a liquid; see Ref.\ \cite{vanMeel09b} for a detailed discussion.
Below, we will express many quantities in dimensionless (LJ) units.

\begin{figure}[htbp]
\includegraphics[width=2.9in]{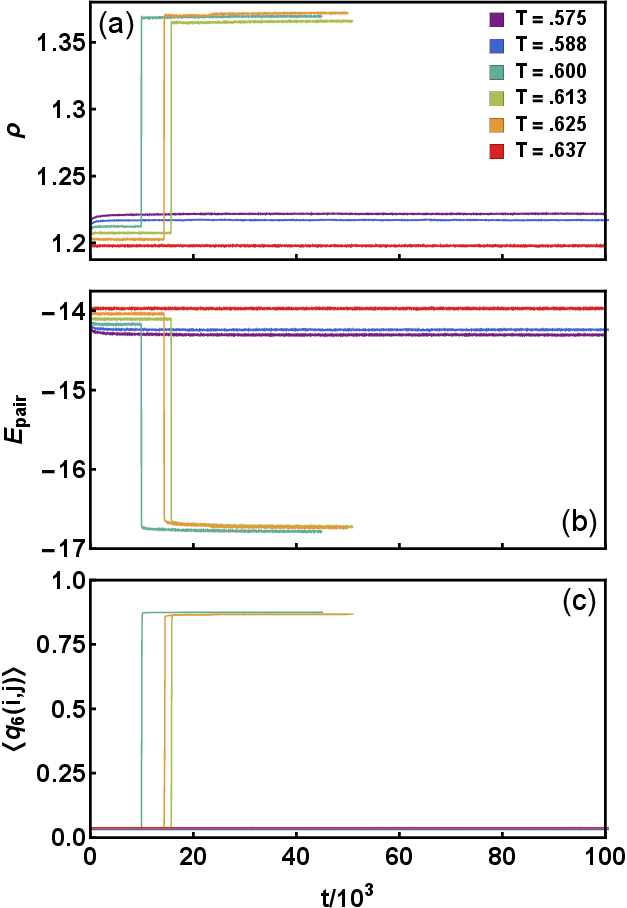}
\caption{Number density $\rho$, average pair energy $E_{\rm pair}$, and average  two-particle bond-order correlator $\langle q_6(i,j) \rangle$ in supercooled 4D Lennard-Jones liquids maintained at zero pressure and the temperatures indicated in the legend of panel (a).}
\label{fig:1}
\end{figure}

\section{Results}
\label{sec:results}

Figure \ref{fig:1} illustrates the evolution of three structural metrics at various temperatures $0.575 \leq T \leq 0.6375$.
In all systems with $0.600 \leq T \leq 0.625$, $E_{\rm pair}$ drops sharply while $\rho$ and $\langle q_6(i,j) \rangle$ increase sharply at various $t < 2\times 10^4$; here $t$ is the time elapsed since the beginning of the NPT runs, and  the average is taken over all $(i,j) \in [1,N]$.
In systems with with lower and higher $T$, including additional values outside the range shown here, no such rapid changes in any of these quantities occur.
These results indicate that the range of temperatures over which the 4D Lennard-Jones liquid crystallizes fastest (at zero pressure) is $0.59 \lesssim T \lesssim 0.63$.
Note that our $N = 5\times 10^5$ systems had periodic simulation cell side lengths $L \gtrsim 25\sigma$ for all $t$, removing the possibility that crystallization was promoted by the periodic boundary conditions as may have been the case \cite{foot648} for the 648-atom systems studied in Ref.\ \cite{skoge06}.

To better understand the role played by the attractive interactions, we followed the strategy employed in Ref.\ \cite{toxvaerd21} and performed NVT runs of length $2.5\times 10^5\tau$ for WCA liquids at $T = 0.600,0.613,0.625$, all at the same $\rho$ values the corresponding LJ liquids had just before they crystallized.  
None of these systems showed any signs of crystallization; in all three cases,  $\langle q_6(i,j)(t) \rangle$ remained stable at its liquid-state value ($\sim 0.03$).

\begin{figure}[h!]
\includegraphics[width=3in]{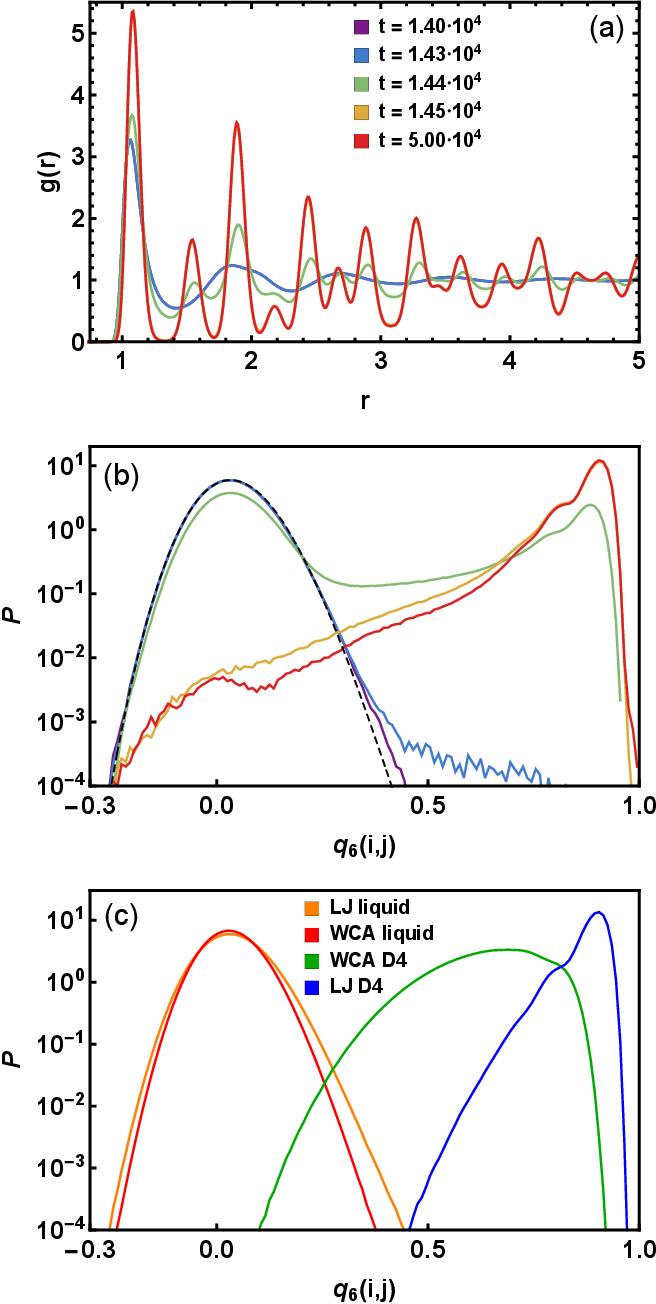}
\caption{Structural order in 4D Lennard-Jones/WCA systems at $T=0.625$. 
Panels (a-b) respectively show the pair correlation functions $g(r)$ and the probability distributions $P[q_6(i,j)]$ for Lennard-Jones systems at selected times, while panel (c) compares time-averaged results for LJ and WCA liquids at the same $\rho = 1.202$ as well as LJ and WCA single $D_4$ crystals at the same $\rho = 1.372$.  The dashed curve in panel (b)  shows a fit to Eq.\ \ref{eq:pliq}, with $a = 113$, $b = 115$, $c = 25$, and $\hat{q}_6 = .031$.} 
\label{fig:2}
\end{figure}

Next we examined the evolution of the LJ systems' structural order.
Results for a single representative temperature ($T = 0.625$) are shown in Figure \ref{fig:2}.
Panel (a) focuses on the pair correlation function $g(r)$.
At $t = 1.40\times 10^4$ and $1.43\times 10^4$, $g(r)$ takes a typical liquid-state form.
At $t = 1.44\times 10^4$, the system is evidently in an intermediate state containing (unstably) coexisting liquid and crystalline regions.
At $t = 1.45\times 10^4$ and all later times, e.g.\ $t = 5.00\times 10^4$ as shown in the plot, the system has solidified into a high-quality crystal.
We verified that it is in fact a $D_4$ crystal by checking that the coordination number $Z = \int_0^{1.4} 2\pi^2 \rho r^3 g(r) dr$ [the 4D analogue of the familiar three-dimensional formula $Z = \int_0^{1.4} 4\pi \rho r^2 g(r) dr$] is very close to $24$.
Specifically, $Z \simeq 23.81$ at  $t = 1.45\times 10^4$, and thereafter continues increasing slowly with $t$; the observation of $Z \to 24$ rules out the competing ($A_4$) crystal structure \cite{vanMeel09b}.

As expected from the decorrelation principle \cite{torquato06,skoge06}, the secondary and tertiary peaks of the liquid-state $g(r)$ are slightly less sharp than in a comparable metastable 3D liquid \cite{toxvaerd21}.
On the other hand, as expected from previous studies of four-, five-, and six-dimensional hard-sphere systems  \cite{lue10,lue21}, the coordination shells of the $D_4$ crystal are much more sharply defined those of the corresponding 3D crystal. 
For example, if $r_{\rm max}$ and $r_{\rm min}$ correspond to the first maximum and minimum of $g(r)$, the $D_4$ crystal has $g(r_{\rm max}) \simeq 5.5$ and $g(r_{\rm min}) \simeq 0.013$ whereas a comparable FCC crystal has $g(r_{\rm max}) \simeq 4$ and $g(r_{\rm min}) \simeq 0.2$ \cite{toxvaerd21}.
This combination of less-correlated liquid structure and more-correlated crystal structure substantially increases entropic contributions to the free energy barriers for crystal nucleation \cite{skoge06,vanMeel09,vanMeel09b,lue10,charbonneau13}.

Why, then, do our 4D LJ systems form high-quality crystals when their hard-sphere and WCA counterparts do not?
Further insight can be obtained by looking at the probability distributions for $q_6(i,j)$. 
As illustrated in Fig.\ \ref{fig:2}(b), $P[q_6(i,j)]$ in the liquid state is well fit by the positively-skewed-Gaussian distribution
\begin{equation}
\footnotesize
P_{\rm liq}(q_6) = P_{\rm liq}(\hat{q}_6)\exp\left[ -a(q_6 - \hat{q}_6)^2 + b(q_6 - \hat{q}_6)^3  - c(q_6 - \hat{q}_6)^4 \right], 
\label{eq:pliq}
\end{equation}
where $\hat{q}_6$ is the mode (most likely) $q_6$ value.
$a$, $b$, $c$ and $\hat{q}_6$ are all positive; both the positive $\hat{q}_6$ and the positive $b$ (i.e.\ the skew of the distribution towards $q_6 > \hat{q}_6$) arise because higher-$q_6$ configurations have lower $E_{\rm pair}$.
Crystal nucleation is indicated by the appearance of a high-$q_6$ tail in $P[q_6(i,j)]$ at $t = 1.43\cdot 10^4$.
At this point, local bond order in the system is $\sim 0.014\%$ crystalline as measured by the fraction of ($i,j$) pairs that have $q_6(i,j) \geq 0.4$ \cite{vanMeel09}.
This fraction grows to $\sim 33\%$ over the next $100\tau$, and by $t = 1.45\cdot 10^4$ it is $99.2\%$.
Thereafter it continues growing slowly as defects gradually anneal out.

Next we compared these time-dependent $P[q_6(i,j)]$ results to time-averaged $P[q_6(i,j)]$ for Lennard-Jones and WCA liquids and $D_4$ single crystals \cite{conway93} at the same temperatures and densities.
The single-crystal systems were prepared using NVT equilibration runs at the densities the LJ systems reached in the limit  $t\to\infty$, e.g.\ $\rho = 1.372$ for $T = 0.625$  [Fig.\ 1(a)].
As illustrated in Fig.\ \ref{fig:2}(c), the LJ liquids' $P[q_6(i,j)]$ are broader and have much longer high-$q_6$ tails compared to the WCA liquids, while the WCA \textit{crystals}' $P[q_6(i,j)]$ are much broader and have much longer \textit{low}-$q_6$ tails compared to the LJ crystals.
Both of these trends are also present  in 3D systems \cite{toxvaerd21}.

A key difference of our results from those of Ref.\ \cite{toxvaerd21}, however, is that in 4D systems the latter trend is much stronger.
As a consequence, $\mathcal{O}$ (Eq.\ \ref{eq:OQ6}) is about four orders of magnitude larger for the WCA systems than it is for the LJ systems (Table \ref{tab:OandDEpair}).
Larger $\mathcal{O}$ lower the entropic barriers to crystallization by increasing the likelihood for localized regions within a supercooled liquid to have crystal-like bond order \cite{toxvaerd20}; this was cited as the primary reason why 3D LJ systems crystallize much faster than their WCA counterparts \cite{toxvaerd21}.
In 4D, however, it appears that LJ systems' much larger $-\Delta E_{\rm pair}$ more than makes up for their much smaller $\mathcal{O}$.
In other words, it appears that energy trumps entropy in determining how rapidly 4D WCA/LJ systems crystallize.

\begin{table}[htbp]
\caption{$\Delta E_{\rm pair} =  E_{\rm pair}^{cryst} - E_{\rm pair}^{\rm liquid}$ and $\mathcal{O}$ for the three $T$ for which LJ liquids crystallized.}
\begin{ruledtabular}
\begin{tabular}{lccc} 
Potential & $T$ & $\Delta E_{\rm pair}$ & $\mathcal{O}$ \\
WCA & 0.600 & -0.356 & $1.4\times10^{-4}$\\
 & 0.613 & -0.356 &  $1.4\times10^{-4}$\\
 & 0.625 & -0.356 &  $1.5\times10^{-4}$\\
 LJ & 0.600 & -2.751 & $3.1\times10^{-9}$\\
 & 0.613 &  -2.807 & $2.1\times10^{-9}$ \\
 & 0.625 & -2.866 &  $2.0\times10^{-9}$\\
\end{tabular}
\end{ruledtabular}
\label{tab:OandDEpair}
\end{table}

\section{Discussion and Conclusions}
\label{sec:conclude}

 Refs.\ \cite{vanMeel09,vanMeel09b} showed that the absence of the most obvious type of geometrical frustration, namely the incompatibility of the lowest-energy/maximally-dense local structure with the ground-state crystal, does not mean that supercooled 4D hard-sphere liquids can easily crystallize.
This is true because the \textit{actual} local structure of these liquids is very different than that of  their maximally-dense local structure.
The overlap $\mathcal{O}$ (Eq.\ \ref{eq:OQ6}) of local bond order distributions in the supercooled-liquid and equilibrium-crystalline states is surprisingly (and substantially) less in 4D than it is in 3D, and it continues to decrease rapidly, leading to free energy barriers to crystallization that grow rapidly, with increasing $d$  \cite{vanMeel09,vanMeel09b}.
This trend is consistent with the recent observation that increasing $r_c$ in 3D WCA/LJ systems substantially increases both their $\mathcal{O}$ values and their nucleation rates \cite{toxvaerd21}.

Here we showed that 4D WCA/LJ systems violate this paradigm.
Specifically, we showed that supercooled $N = 5\times10^5$ LJ liquids maintained at zero pressure and constant temperatures $0.59 < T < 0.63$ formed high-quality $D_4$ crystals within $\sim 2 \times 10^4\tau$, whereas WCA liquids that were maintained at the same densities and temperatures at which their LJ counterparts nucleated 
did not crystallize even after $2.5\times 10^5\tau$, despite the fact that the WCA systems had $\mathcal{O}$ values that were several orders of magnitude larger.
One could certainly have expected that the LJ systems' much larger $-\Delta E_{\rm pair}$ (Table \ref{tab:OandDEpair}) would dramatically speed up their crystallization, but ours was the first actual observation of homogeneous crystallization in simulated $d > 3$ liquids that followed physical (Newtonian) dynamics \cite{biasedMC} and were large enough to eliminate the possibility that crystallization was promoted by the periodic boundary conditions.
It was enabled by \texttt{hdMD}'s efficient parallel implementation, which allowed us to perform simulations with an $N$ that was at least ten times larger than those employed in any previous $d > 3$ simulations other than those of Ref.\ \cite{hoy22}.

Since $-\Delta  E_{\rm pair}$ in systems with at-least-intermediate-range attractive interactions probably continues to grow with $d$ \cite{lue10,lue21,charbonneau21}, our results suggest that the widely-accepted hypothesis that crystallization rapidly gets harder with increasing $d$ \cite{skoge06,vanMeel09,vanMeel09b}  is only generally valid \textit{in the absence of attractive interparticle forces.}
At the very least, when considered in combination with our demonstration that higher $\mathcal{O}$ values do not necessarily lead to higher nucleation rates, they suggest that accounting for energetic (not just entropic) contributions to the free energy barriers to crystallization is necessary to determine whether this hypothesis is true in general.

We conclude by mentioning one additional potential implication of our results for future studies of the glass transition.
Monodisperse 4D hard-sphere and repulsive-soft-sphere liquids are particularly useful for studies of this transition \cite{charbonneau10,charbonneau13,eaves09,charbonneau13b} because they lack the strong correlations between particle size and particle mobility  which make it challenging to interpret the heterogeneous dynamics of model  bidisperse and polydisperse supercooled liquids \cite{kob95,kob97,donati98}.
Our results demonstrate another way such liquids can be useful: their crystallization propensity can be tuned by varying their cutoff radius $r_c$.

We thank Patrick Charbonneau for helpful discussions.
This material is based upon work supported by the National Science Foundation under Grant DMR-2026271.


\begin{thebibliography}{42}
\bibitem [{foo()}]{foot648}%
  \BibitemOpen
  \href@noop {} {}\bibinfo {note} {Since the $D_4$ lattice has 8 atoms per unit
  cell \cite{conway93}, the choice $N = 648$ promotes the formation of a
  $3\times3\times3\times3$-cell $D_4$ single crystal.}\BibitemShut {Stop}%
\bibitem [{bia()}]{biasedMC}%
  \BibitemOpen
  \href@noop {} {}\bibinfo {note} {Homogeneous crystallization of larger 4D
  hard-sphere liquids has been induced using configurational-bias Monte Carlo
  \cite{vanMeel09,vanMeel09b}, but it has not yet been reported in systems
  following the normal physical dynamics.}\BibitemShut {Stop}%
\end{thebibliography}

%

\end{document}